\documentclass[%
 reprint,
 superscriptaddress,
 amsmath,amssymb,
 aps,
 pra,
]{revtex4-2}

\usepackage{graphicx}
\usepackage{dcolumn}
\usepackage{bm}
\usepackage{braket}
\usepackage{hyperref}

\usepackage{booktabs}
\usepackage{xcolor}
\usepackage{siunitx,xfrac}
\DeclareSIUnit{\dBm}{dBm}
\DeclareSIUnit{\dBc}{dBc}
\DeclareSIUnit{\dBi}{dBi}
\DeclareSIUnit{\Vcm}{(V/cm)}
\DeclareSIUnit{\cmV}{(cm/V)}
\DeclareSIUnit{\uVcm}{(\micro V/cm)}
\DeclareSIUnit{\uVm}{(\micro V/m)}
\DeclareSIUnit{\Vm}{(V/m)}
\DeclareSIUnit{\mVm}{(mV/m)}

\renewcommand{\Re}{\operatorname{Re}}
\renewcommand{\Im}{\operatorname{Im}}

\newcommand{\OD}{\operatorname{OD}}

\begin{document}

\preprint{APS/123-QED}

\title{Optimal Atomic Quantum Sensing using EIT Readout}

\author{David H. Meyer}%
 \email{Corresponding author: david.h.meyer3.civ@army.mil}
 \affiliation{DEVCOM, Army Research Laboratory, Adelphi MD 20783, USA}
\author{Christopher O'Brien}
 \affiliation{Naval Air Warfare Center, China Lake CA 93555, USA}
\author{Donald P. Fahey}
 \affiliation{DEVCOM, Army Research Laboratory, Adelphi MD 20783, USA}
\author{Kevin C. Cox}
 \affiliation{DEVCOM, Army Research Laboratory, Adelphi MD 20783, USA}
\author{Paul D. Kunz}
 \affiliation{DEVCOM, Army Research Laboratory, Adelphi MD 20783, USA}

\date{\today}

\begin{abstract}
Quantum sensors offer the capability to reach unprecedented precision by operating at the standard quantum limit (SQL) or beyond by using quantum entanglement.
But an emerging class of quantum sensors that use Rydberg electromagnetically-induced transparency (EIT) to detect rf electric fields have yet to reach the SQL.  In this work we prove that this discrepancy is due to fundamental limitations in the EIT probing mechanism.  We derive the optimum sensitivity of a three-level quantum sensor based on EIT, or more generally coherent spectroscopy, and compare this to the SQL.  We apply a minimal set of assumptions, while allowing  strong probing fields, thermal broadening, and large optical depth. We derive the optimal laser intensities and optical depth, providing specific guidelines for sensitive operation under common experimental conditions. Clear boundaries of performance are established, revealing that ladder-EIT cannot achieve the SQL due to unavoidable absorption loss. The results may be applied to any EIT-based quantum sensor, but we particularly emphasize our results' importance to the growing field of Rydberg quantum sensing.
\end{abstract}

\maketitle

\section{Introduction}
Quantum sensors based on atomic vapors offer state-of-the-art performance across a variety of precision measurement applications, from clocks \cite{ludlow_optical_2015} to magnetometers \cite{budker_resonant_2002,budker_optical_2007} to inertial sensors \cite{gauguet_characterization_2009,muller_atom-interferometry_2008,cronin_optics_2009}. A relatively new quantum sensor, based on Rydberg atoms measuring electric fields \cite{mohapatra_giant_2008, sedlacek_microwave_2012}, has already demonstrated record-breaking precision over traditional classical methods \cite{holloway_electric_2017, jing_atomic_2020}, yet is still far from reaching the limits of its optimal sensitivity. Electromagnetically induced transparency (EIT) is ubiquitously used in nearly all Rydberg sensors, and indeed many other quantum sensors as well. Here we take an essential step forward, showing how to optimize EIT for quantum sensing, with particular attention to Rydberg sensors, and reveal fundamental limitations of EIT that prevent achievement of the standard quantum limit. We hope this work serves two primary purposes: 1) give optimal parameters for improving current-generation EIT sensors, 2) inspire investigation of alternate methods that do reach the standard quantum limit, obtaining maximal information from a given number of non-entangled particles.

Recent advances in Rydberg quantum sensor research from groups around the world are leading to rapid development \cite{holloway_broadband_2014,fan_atom_2015,cox_quantum-limited_2018, deb_radio-over-fiber_2018, meyer_assessment_2020, meyer_digital_2018, shaffer_read-out_2018, thaicharoen_electromagnetically_2019, jau_vapor-cell-based_2020, jing_atomic_2020}. In addition to precise measurement of weak \cite{jing_atomic_2020} and strong fields \cite{paradis_atomic_2019}, these sensors offer operating frequency range spanning from quasi-DC \cite{jau_vapor-cell-based_2020} to the terahertz regime \cite{downes_full-field_2020}, sub-wavelength imaging \cite{holloway_sub-wavelength_2014,wade_real-time_2017}, and reception of AM, FM, and Wi-Fi broadcast signals \cite{meyer_waveguide-coupled_2021}. 
These sensors are poised to have a broad impact across society analogous to the impact of
vapor quantum sensors that are now used in GPS satellites, magnetoencephalography \cite{boto_moving_2018}, and geodesy/navigation \cite{bongs_taking_2019}. However, in order for these sensors to achieve broad impact, improvements are still necessary, particularly in the sensitivity \cite{meyer_assessment_2020}.

After providing some background and introducing our theoretical model, we first derive an analytic solution for the optimum sensitivity of EIT quantum sensors for stationary atoms in an optically-thin ensemble. This gives insights to guide our numerical models of more complex situations that include large optical depths and Doppler effects. Although signal-to-noise ratio often improves as the square-root of atom number, as optically-thick samples are considered there is unavoidable absorption loss associated with EIT. We find that ladder-EIT is strictly worse than the standard quantum limit (SQL) by a factor of approximately \SI{3.5}{\deci\bel}, while lambda-EIT can asymptotically reach the SQL for high atom numbers.  Finally, we introduce Doppler-broadening due to warm atoms and calculate the optimum sensitivity in the ladder scheme, showing that the limiting value of \SI{3.5}{\deci\bel} may be recovered in the limit of large optical depth and Rabi frequencies. In sum, our results show the theoretical optimum sensitivity of an EIT quantum sensor, give a recipe for sensitive operation, and define a clear boundary for performance.  This work solves the outstanding discrepancy between the sensitivity of state-of-the-art Rydberg sensors and the SQL \cite{fan_atom_2015,meyer_assessment_2020} and serves as a foundation upon which to devise more effective experimental methods.

\section{Background}

The optimal precision for any quantum sensor when relying on uncorrelated or non-entangled states is set by the number of quantum particles involved in the measurement, and is known as the quantum projection limit or standard quantum limit (SQL) \cite{itano_quantum_1993}. Achieving this optimum is challenging, but techniques such as Ramsey spectroscopy are known to be ideal methods \cite{wineland_squeezed_1994} and have realized optimal signal-to-noise ratios at the SQL \cite{santarelli_quantum_1999}. An alternate technique, electromagnetically-induced-transparency (EIT)
\footnote{We categorize electromagnetically-induced-transparency (EIT) \cite{fleischhauer_electromagnetically_2005}, coherent population trapping (CPT) \cite{arimondo_v_1996}, and nonlinear magneto-optical rotation (NMOR) \cite{budker_resonant_2002} all as specific regimes of coherent spectroscopy. We adopt the broader usage of the term EIT to encompass the methods studied here in order to connect with relevant communities that use this term widely.},
which involves coherent dark states established using three (or more) energy levels, is one of the most commonly used in quantum sensing. 
In particular, current implementations of the Rydberg atomic electric field sensor rely on EIT spectroscopy, where the long-lived, high-energy Rydberg state and ground state form the coherent dark state \cite{mohapatra_coherent_2007}. 
While the SQL sensitivity of these sensors, derived from quantum projection noise, is projected to be of order \SI{10}{\pico\volt\per\centi\meter\sqrt{\hertz}} \cite{fan_atom_2015}, the current record performance falls far short of the SQL \cite{jing_atomic_2020,jau_vapor-cell-based_2020}. Previous estimates of the SQL performance did not fully account for the EIT probing process, which can be safely assumed to have an implicit quantum efficiency of detection that will limit the performance.
In this Article we study the optimal performance of EIT spectroscopy methods relative to the SQL, with particular focus applied to Rydberg ladder-EIT field sensors, in order to provide a more realistic estimate of the effective SQL performance.

EIT and related methods \footnotemark[\value{footnote}] have been extensively studied in many contexts beyond sensing. For example, optical quantum memories, slow and fast light, and lasing without inversion have received much attention \cite{fleischhauer_electromagnetically_2005}, and our work builds on this broad foundation. In most previous work there are several common assumptions such as weak-probing, optically-thin samples, or no Doppler-broadening; here we will go beyond each of these assumptions. 

Previous work in the context of EIT-magnetometers, which rely on the so-called lambda-configuration,  has shown that the weak-probe regime is not optimal \cite{fleischhauer_quantum_2000}, but did not consider optically-thick samples and the effect of Doppler-broadening was negligible. Other magnetometry work has considered optically-thick Doppler-broadened samples \cite{rochester_nonlinear_2002}, but did not explicitly compare to the SQL.  We derive the optimum sensitivity of an EIT-based quantum sensor using a semi-classical model with quantum noise in the detection process.
We apply minimal assumptions, and show that optimal performance naturally requires operating with both high optical depth and strong optical fields. We note that additional effects not considered here, such as back-action from AC-Stark shifts \cite{fleischhauer_quantum_2000}, radiation trapping \cite{matsko_radiation_2001}, and non-uniform Gaussian beam profiles could likely place even more stringent limitations on the sensitivity of realistic devices.

The sensitivity benchmark that we use throughout this work is that of an ideal quantum sensor operating at the standard quantum limit. We model this system by taking an ensemble of $N$ independent two-level atoms.  The sensor detects an ambient field $F$ by observing an energy shift $\Delta E = \hbar\, \delta$ on the excited state, given by $\hbar\, \delta = d\,F$ where $\hbar$ is the reduced Planck constant and $d$ is a discriminator that characterizes the interaction strength between the atoms and $F$. Throughout this work we restrict our analysis to the sensor's ability to measure perturbations in $\delta$. Doing so allows our results to be applied to any measurement of a weak field $F$ via a simple conversion factor defined by the discrimination coefficient $d$. See Appendix \ref{app:FieldSens} for further details.

A prototypical quantum sensor of two-level atoms is initialized into a non-entangled superposition state that is characterized by quantum phase $\phi$.  The phase evolves in time $\phi=\delta\,T$ during a measurement time $T$ (often visualized as a rotating vector around a Bloch sphere). The theoretical optimum precision with which $\phi$ can be measured is limited by quantum projection noise in the readout process, $\Delta\widetilde{\phi}_{SQL}=1/\sqrt{N}$. In the presence of decoherence, where the coherence time of the atoms $T_2$ is shorter than the measurement time, the standard quantum limited minimum detectable shift is given by
\begin{equation}\label{eq:SQL}
    \delta_{min}=\sqrt{\frac{2e}{N T_2 T}}
\end{equation}
found by setting the signal-to-noise ratio (SNR) to one, $\phi/\Delta\widetilde{\phi}=1$, and assuming an optimal Ramsey measurement \cite{huelga_improvement_1997}, where $e$ is the base of the natural logarithm. In this work we will compare this limit to the theoretical optimum performance of spectroscopic readout of $\delta$ via ensembles of 3-level atoms.

\begin{figure}[t]
    \centering
    \includegraphics{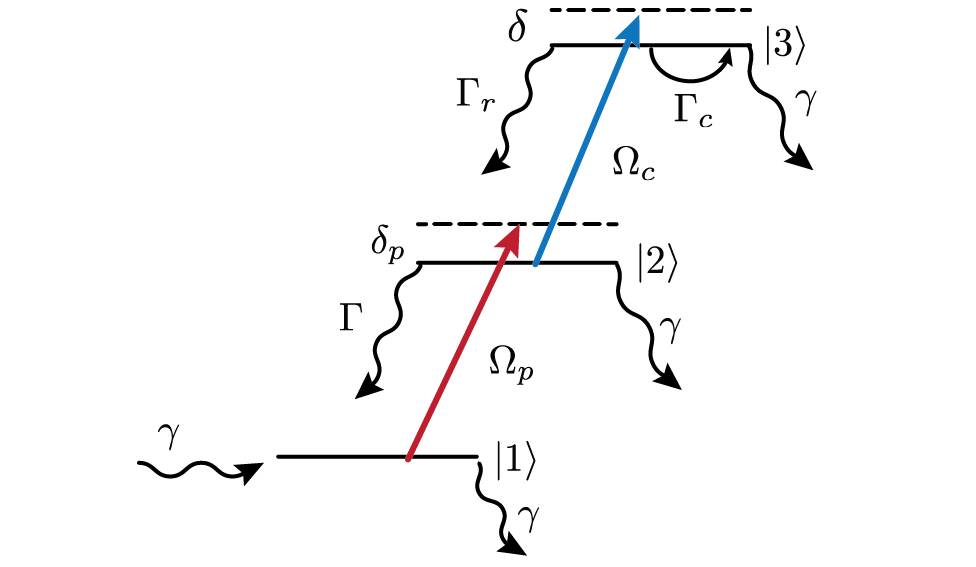}
    \caption{Level diagram for the 3-level ladder scheme common to Rydberg-EIT sensors. The probing (coupling) optical field $\Omega_p$ ($\Omega_c$) is tuned near resonance with detuning $\delta_p$ ($\delta$). Dephasing processes include natural lifetimes of the intermediate probing state $\Gamma$ and excited state $\Gamma_r$, collisional dephasing of the excited state $\Gamma_c$, and transit dephasing $\gamma$ on all states.}
    \label{fig:levelDiagram}
\end{figure}

\section{Model Definition}

We model an ensemble of $N$ 3-level atoms coupled to two light fields using a standard, semi-classical density matrix approach with the master equation \cite{fleischhauer_electromagnetically_2005}
\begin{equation}\label{eq:masterEq}
    \frac{d\rho}{dt} = -\frac{i}{\hbar}\left[H,\rho\right] + L
\end{equation}
where $H$ is the system Hamiltonian in the rotating frame with counter-rotating terms dropped, $\rho$ is the density matrix, and $L$ is the Lindbladian operator that defines the dephasing terms; including natural lifetimes of the intermediate, $\Gamma$, and excited, $\Gamma_r$, states, collisional dephasing of the excited state $\Gamma_c$, and transit dephasing $\gamma$ resulting from atoms transiting the cross-section of the optical beams (diagrammed in Fig.~\ref{fig:levelDiagram}, see App.~\ref{app:Model} for details).  We treat the optical fields semi-classically, specifically the limitations to the sensor due to quantum noise in the detection process of the probe field.  This approximation is valid for low sensing bandwidths in the steady-state regime.  Further work may be necessary to incorporate additional sensor limitations due to deleterious quantum back-action that occurs at high optical depths and optical powers \cite{fleischhauer_quantum_2000,matsko_radiation_2001}. However, even the relatively simplified treatment given here is sufficient to bound EIT-based quantum sensors at a level more stringent than the SQL.  

The presence of the atomic ensemble causes absorption and phase shifts on the probe light that depend on the density matrix via the differential equations \cite{*[{See Chapter 10 of }] [{. Note this reference uses CGS units.}] auzinsh_optically_2010}
\begin{align}
    \frac{d\Omega_p}{dz} &= \kappa \Im(\rho_{12})\label{eq:diffEq-rabi}\\
    \frac{d\phi}{dz} &= \frac{\kappa}{\Omega_p(z)} \Re(\rho_{12})\label{eq:diffEq-phase}
\end{align}
where $\rho_{12}$ is the density matrix element that describes the atomic coherence between the ground and first excited state and is found by solving Eq.~\ref{eq:masterEq} in steady-state. The prefactor is a constant defined as
\begin{equation}\label{eq:kappa}
    \kappa = \frac{\omega n \mu^2}{2c\epsilon_0 \hbar}
\end{equation}
where $\omega$ is the probing transition frequency, $n$ is the atomic number density, $\mu$ is the dipole moment of the probing transition, $c$ is the speed of light, $\epsilon_0$ is the dielectric constant, and $\hbar$ is the reduced Planck constant.

By applying the appropriate probe and coupling light fields to the atomic ensemble, we can observe spectroscopic features that depend on the coupling laser detuning, $\delta$, allowing us to perform a measurement of any Stark shift applied to state $\ket{3}$ via an external perturbation. We model an ideal measurement by assuming a photon-shot-noise (PSN) limited homodyne measurement of the acquired phase $\phi$ on the transmitted probe light, where $\phi$ is found by solving the coupled differential Eqs.~\ref{eq:diffEq-rabi} and \ref{eq:diffEq-phase}. The noise density $\Delta\widetilde{\phi}$ in measuring this quantity for a coherent state is defined as
\begin{equation}\label{eq:phase-noise}
    \Delta\widetilde{\phi}=\frac{1}{\sqrt{4\dot{M}_d}}=\frac{\eta}{\Omega_p(\ell)}
\end{equation}
where $\dot{M}_d$ is the detected photon number flux, $\Omega_p(\ell)$ is the transmitted probe Rabi frequency, and $\eta$ is defined as
\begin{equation}
    \eta = \sqrt{\frac{\omega\mu^2}{2qc\epsilon_0\hbar A}}
\end{equation}
where $q$ is the detection efficiency (assumed to be unity in the remainder of this work), and $A$ is the cross-sectional area of the probing field. 

Further details regarding these definitions are found in Appendix \ref{app:Model}.

\section{Calculating Sensitivity}
\subsection{Optically-Thin Ensemble}
To gain intuition, we begin with the simple case of a non-Doppler broadened, optically-thin sample and find an analytic solution. In this approximation, the sensor signal-to-noise density is given by
\begin{equation}\label{eq:SNR}
    \frac{\phi(\delta)}{\Delta\widetilde{\phi}} =\frac{\kappa  \delta \ell}{\eta} \frac{d}{d\delta} \Re(\rho_{12}) 
\end{equation}
where $\ell$ is the optical path length. 
The maximum slope of $\Re(\rho_{12})$ occurs at EIT resonance ($\delta_p=\delta=0$), and we approximate it by setting the single-photon detuning $\delta_p$ to zero and taking only the first-order term in small perturbations of the Rydberg state detuning $\delta$ around zero. The associated minimum detectable shift in one second $\widetilde{\delta}_{min}$ \footnote{We denote densities (versus absolute values) with the over tilde. Converting an absolute minimum, like that defined in Eq.~\ref{eq:SQL}, to a density requires setting $T=\SI{1}{\second}$.}, or sensitivity, is
\begin{equation}
    \widetilde{\delta}_{min} = \frac{\eta}{\kappa\ell}\left(\frac{d}{d\delta}\Re(\rho_{12})\right)^{-1}.
\end{equation}

\begin{figure}[t]
    \centering
    \includegraphics[width=\columnwidth]{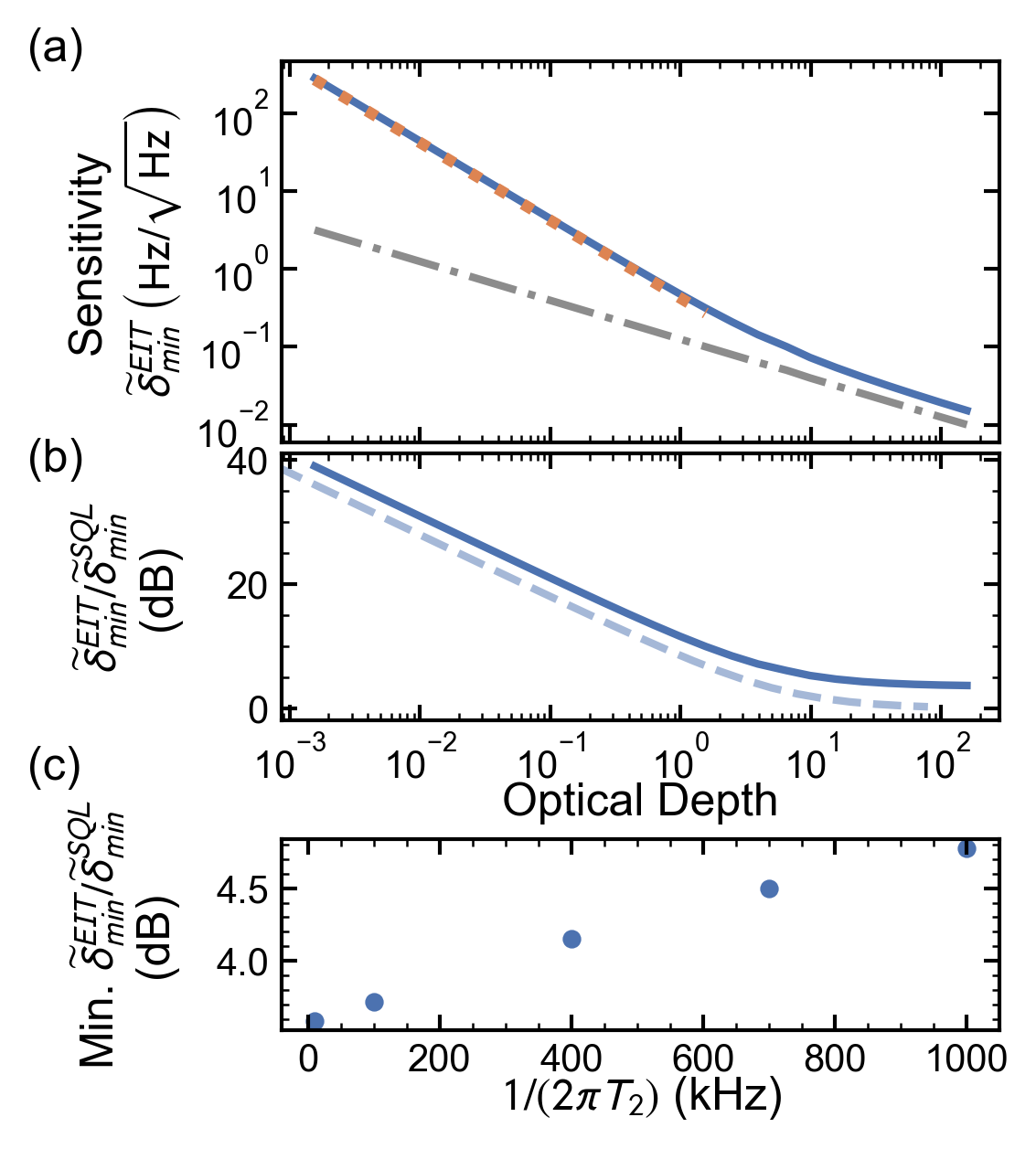}
    \caption{Sensitivity of optically-thin and thick ensembles.
    (a) Sensitivity with optimized probe and coupling Rabi frequencies versus optical depth (solid blue line). The dotted orange line is the analytic result from Eq.~\ref{eq:AnalyticOpt}. The gray dash-dot line is the SQL result from Eq.~\ref{eq:SQL}.
    Parameters: $\gamma=2\pi\times\SI{100}{\kilo\hertz}$, $\Gamma_r=2\pi\times\SI{1}{\kilo\hertz}$.
    (b) The ratio of optimal EIT sensitivity to the SQL, expressed in dB of variance. At the optimal optical depth, EIT probing is approximately \SI{3.5}{\deci\bel} worse than the SQL. The light blue dashed line is the same ratio for a corresponding lambda-configuration, which asymptotes to the SQL (\emph{i.e.} $\SI{\sim0}{\deci\bel}$).
    (c) Minimum ratio of EIT and SQL sensitivities as a function of $\gamma$ at $\OD\approx100$.}
    \label{fig:OptThick}
\end{figure}

In the idealized case where $\Gamma$ is the only non-zero dephasing rate, the slope in Eq.~\ref{eq:SNR} reduces to
\begin{equation}\label{eq:noDephaseSlope}
    \frac{d}{d\delta}\Re(\rho_{12})\approx\frac{2\Omega_p\Omega_c^2}{(\Omega_p^2+\Omega_c^2)^2}.
\end{equation}
The ratio of Rabi frequencies that optimizes this slope is found to be $\Omega_c=\sqrt{2}\Omega_p$ (see Appendix \ref{app:OptimalRabi} for details).
This is an important result that remains approximately true for small dephasing rates and indicates that optimal sensitivity for optically-thin media with a single probing beam does not occur in the weak probe EIT regime where $\Omega_p\ll\Omega_c$ nor when $\Omega_p=\Omega_c$. It also extends to other EIT configurations, such as the lambda scheme, since the intermediate state is adiabatically eliminated from the steady-state solution in the limit of small coherent state dephasing.

The resulting optimal sensitivity is then
\begin{equation}
    \widetilde{\delta}_{min} = \frac{\eta}{\kappa\ell}\frac{9\Omega_p}{4}.
\end{equation}
From this we see the sensitivity is improved ($\widetilde{\delta}_{min}$ reduced) for small Rabi frequencies. Note however, that the width of the EIT resonance, and therefore the regime of linear phase accumulation near $\delta=0$ is also made arbitrarily small (scaling as approximately $3\Omega^2/\Gamma$). 

If we take this ratio as an ansatz, we can derive the optimal Rabi frequencies and sensitivity when dephasing is present to first order in the dephasings ($\gamma,\,\Gamma_r,\,\Gamma_c\ll\Gamma$).
The optimal Rabi frequencies take the form of a geometric average of the dephasing rates
\begin{align}\label{eq:AnalyticRabis}
    \Omega_p^{(opt)} &\approx\sqrt{\Gamma(2\gamma+\Gamma_r+\Gamma_c)},\\
    \Omega_c^{(opt)} &\approx\sqrt{2\Gamma(2\gamma+\Gamma_r+\Gamma_c)}.
\end{align}
We see that the typical EIT weak probe regime is not optimal, rather the optimum occurs with similar Rabi frequencies between the two optical fields, sometimes known as the Coherent Population Trapping (CPT) regime \cite{arimondo_v_1996}. Never the less, this result is intuitive, as it represents the effective saturation parameter for our three-level ladder system, \emph{i.e.} the system must be pumped at a rate higher than the combined dephasing rate of the coherent state, yet not so high to overly broaden the probing transition. Increasing beyond one saturation parameter begins to hurt the discriminator slope (signal) through power broadening. We also note that the effect of transit broadening is twice that of the Rydberg dephasings since transit also equally affects coherence in the ground state.

The corresponding optimum sensitivity is then
\begin{equation}\label{eq:AnalyticOpt}
    \widetilde{\delta}_{min}^{EIT}\approx \frac{4\eta}{\kappa\ell}\sqrt{\Gamma(2\gamma+\Gamma_r+\Gamma_c)}.
\end{equation}
This result, valid at low optical depth, is plotted as an orange dotted line in Fig.~\ref{fig:OptThick}(a). We see that sensitivity increases linearly with atom number (i.e. larger $\kappa$, cell length, and smaller $\eta$). 

\subsection{Comparison with SQL}

We can directly compare the result from Eq.\ \ref{eq:AnalyticOpt} with the atom projection noise limit in order to determine the extent to which photon shot noise in EIT probing approximates an optimal quantum sensor. Under the assumption that the total dephasing of the coherent state is limiting, the SQL to measuring shifts of the Rydberg state (in one second) reduces to
\begin{equation}\label{eq:SQLOpt}
    \widetilde{\delta}_{min}^{SQL}=\sqrt{\frac{e(2\gamma+\Gamma_r+\Gamma_c)}{n A\ell}}
\end{equation}
where we have taken $N=n A \ell$ to be the total number of atoms in the sensing volume and
\begin{equation}\label{eq:T2}
    T_2=2/(2\gamma+\Gamma_r+\Gamma_c) 
\end{equation}
is the coherence time of the Rydberg state. This result is shown as the dashed gray line in Figure \ref{fig:OptThick}(a).

The ratio of EIT sensitivity to SQL sensitivity is then found to be
\begin{equation}
    \frac{\widetilde{\delta}_{min}^{EIT}}{\widetilde{\delta}_{min}^{SQL}}=\frac{4\eta\sqrt{n A \ell\Gamma}}{\kappa\ell\sqrt{e}}.
\end{equation}

Taking the resonant, small-signal optical depth (OD) to be
\begin{equation}
    \OD=\frac{\omega\mu^2n\ell}{c\epsilon_0\hbar\Gamma}
\end{equation}
we can reduce the ratio of sensitivities to
\begin{equation}
    \frac{\widetilde{\delta}_{min}^{EIT}}{\widetilde{\delta}_{min}^{SQL}}=\frac{4\sqrt{2}}{\sqrt{e\OD}}
\end{equation}
which is valid for $\OD<1$.

The ratio of sensitivities  solely depends on optical depth. As OD increases the EIT photon-shot-noise-limited SNR increases more rapidly than SQL-limited SNR, so the ratio converges; in other words the information from the probe light more closely corresponds to that of the atoms, as expected. This change in scaling is similar to that reported for optical magnetometry \cite{rochester_nonlinear_2002}. At $\OD=0.5$, the EIT sensitivity is a factor of approximately $4.9$ (\SI{13.7}{\deci\bel} in variance) worse than the standard quantum limit. As the OD is increased beyond 1, the assumption that $d\Omega_p/dz=0$ loses validity and absorption losses in the probing field must be considered.

\subsection{Finite Optical Depth}

Optical depth is related to the total atom number of the ensemble and directly influences the total interaction strength between the atoms and the probing light field. To calculate the spectroscopic sensitivity for an optically-thick ensemble, we numerically integrate the coupled differential Eqs.~\ref{eq:diffEq-rabi} and \ref{eq:diffEq-phase} over the length of the extended ensemble. In order to numerically approximate the slope of $\Re(\rho_{12})$ with respect to perturbations from zero of $\delta$, we perform the integration twice for each point with $\delta=\pm\epsilon/2$ where $\epsilon$ is chosen to be suitably small. See Appendix \ref{app:OptimalDepth} for details on how the transmitted probe power and phase accumulation scale versus OD with fixed Rabi frequencies. 

In Figure \ref{fig:OptThick}(a) we show the optimal sensitivity versus length as the blue trace. For each point, we have numerically found the optimal optical Rabi frequencies. We find agreement with the analytic result when $\OD<1$. At higher optical depths, we find that the prevailing scaling with OD (a proxy for atom number) changes to approximately that of the SQL (i.e. $\sqrt{N}$). This change in scaling is also observed in similar optical magnetometry systems for optimized ensemble depth \cite{rochester_nonlinear_2002}. We also find that as the OD is increased the optimal Rabi frequencies transition to being approximately equal (see Appendix \ref{app:OptimalRabi} for details).

The ratio of the optimal sensitivity of EIT to SQL versus length is shown as the blue trace of Figure \ref{fig:OptThick}(b), in terms of \si{\deci\bel} in variance. In the $\sqrt{N}$ scaling region of large OD, the ladder-EIT sensitivity reaches an asymptotic value of approximately $\times1.5$ worse than the SQL sensitivity result (\SI{3.5}{\deci\bel}). We note that the optical depth necessary for this optimal result is, though achievable, fairly large ($\gtrsim10$), and would require specialized experimental design considerations.

We show the corresponding ratio for the lambda-system as a light blue dashed line in Fig.~\ref{fig:OptThick}(b). Only minor changes to the decay pathways and the (thermal) population of the two ground states that make up the dark superposition
are required to convert to a lambda-system; for details see appendix \ref{app:MasterEq}. Because the atoms are roughly equally distributed between the two ground states, the total atom number is twice as large for the same probe OD. Thus to keep the comparison fair, at each OD the two models are compared to respective SQLs of appropriate atom number. The ratio of optimal sensitivity of EIT to SQL follows a similar trend, but is able to asymptotically reach the SQL at high OD.

In Figure \ref{fig:OptThick}(c), we show that the optimal EIT sensitivity and SQL scale similarly with dephasing rate, $T_2$, as can be seen by comparing Eqs.\ \ref{eq:AnalyticOpt} and \ref{eq:SQLOpt}. In Figure \ref{fig:OptThick}(c) the OD is held fixed at approximately 100, corresponding to the largest value modeled. For each point we change $T_2$ by changing $\gamma$ then use the same optimization process from Fig.~\ref{fig:OptThick}(a) to determine the optimal Rabi frequencies and sensitivity. We observe that the ratio of sensitivities has only a weak dependence on $1/T_2$ where a $\times100$ increase in dephasing leads to an increase of only \SI{1.2}{\deci\bel}. This indicates both sensitivities approximately scale as $\widetilde{\delta}_{min} \propto 1 / \sqrt{T_2} $.

We note that in practical implementations, $T_2$ is often coupled to other parameters that complicate this analysis.
For example, while $T_2$ can be independent of OD, as would be the case for simply changing path length, increasing OD by increasing particle density (via heating the cell) can lead to corresponding sensitivity loss through increased radiation trapping, collisional broadening, or other atom-atom interactions. Our simplified model does not fully account for such effects, and therefore they are not explicitly considered here.
In any case, the implication of this result is that a reduction in coherence time $T_2$ is not likely to significantly degrade the sensitivity.

\begin{figure}[t]
    \centering
    \includegraphics[width=\columnwidth]{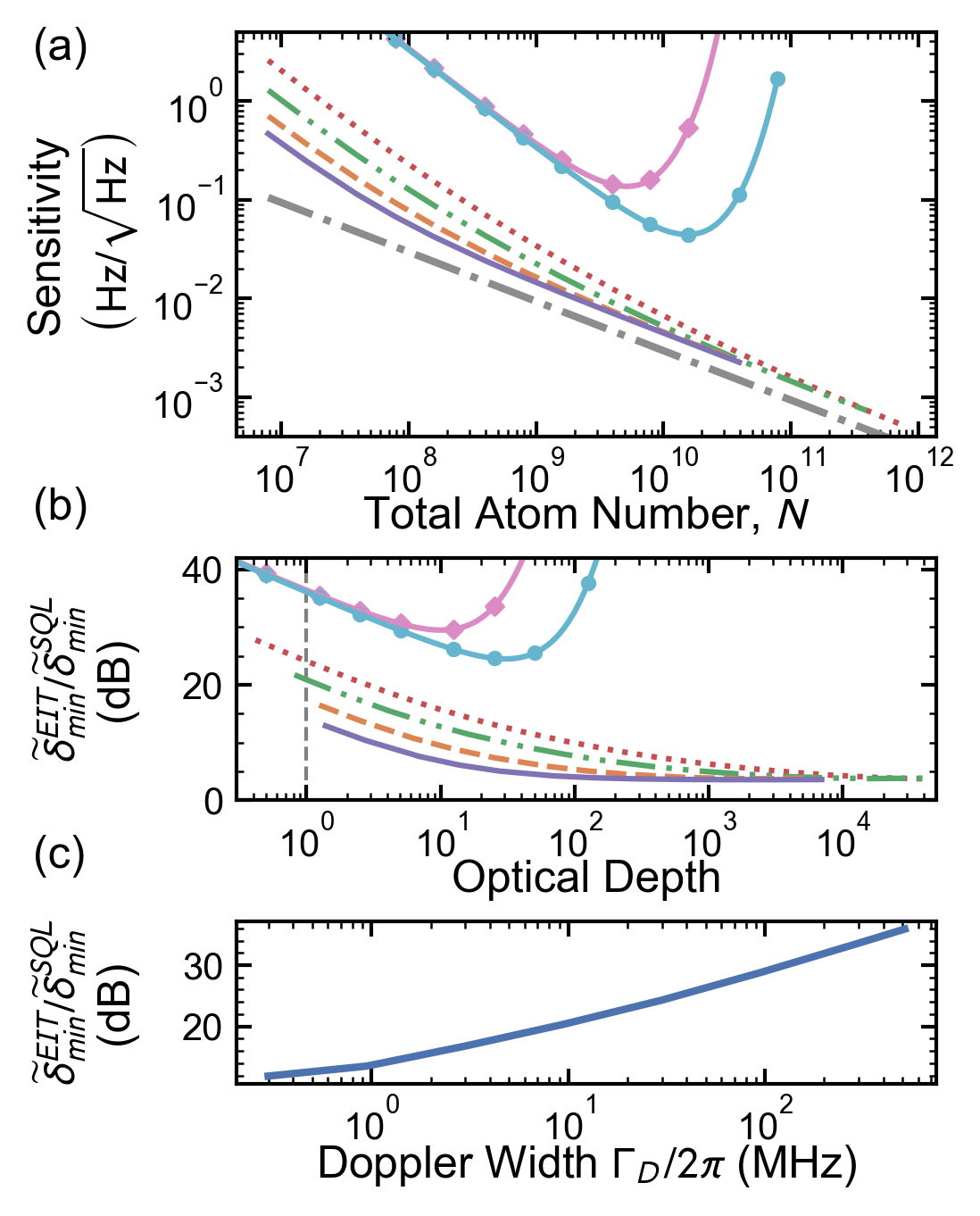}
    \caption{(a) Sensitivity using EIT readout versus total atom number in the sensing volume, $N$. The solid purple, dashed orange,  dash-dot-dot green, and dotted red lines show the optimal sensitivity for an ensemble with Doppler linewidth of 1, 3, 10, and \SI{25}{\mega\hertz} respectively. The gray dash-dot line is the SQL result assuming all atoms in the sensing volume participate equally and the coherence time is set by Eq.~\ref{eq:T2}. The magenta line with diamond points and the cyan line with circle points show the sensitivity with a Doppler width of approximately \SI{520}{\mega\hertz} and non-optimized, fixed $\Omega_p=\Omega_c=2\pi\times 15$ and $\SI{40}{\mega\hertz}$, respectively.
    (b) The ratio of optimum sensitivities and the SQL from part (a).
    (c) Ratio of sensitivities versus Doppler width $\Gamma_D$ at $\OD=1$ (vertical line of part (b)).}
    \label{fig:DopplerLength}
\end{figure}

\subsection{Doppler-Averaging}

Doppler-averaging is the result of atomic motion in an ensemble at finite temperature. Atoms experience Doppler shifts that detune the optical fields from resonance which manifests as a decoherence mechanism.
The inclusion of additional atom velocity classes requires only minor modification of Eqs.~\ref{eq:diffEq-rabi} and \ref{eq:diffEq-phase}. First, the detunings are altered to allow for Doppler shifts: $\delta_i\rightarrow\delta_i\pm k_i v$. Here $k_i=2\pi/\lambda_i$ represents the $k$-vector of the appropriate light field, $v$ the velocity of the atom, and the signs are chosen to be opposite for the probe/coupling field to represent the typical counter-propagating configuration. In the following numeric calculations, we set $\lambda_p=\SI{780.241}{\nano\meter}$ and $\lambda_c=\SI{480.125}{\nano\meter}$, corresponding to a ladder-EIT system in Rubidium-85 formed from the D2 transition and a subsequent transition to a Rydberg state of principal number $n=50$. Second, the matrix element $\rho_{12}$ must be replaced with the Doppler-averaged value
\begin{equation}
    \overline{\rho}_{12}(z)=\frac{1}{v_p \sqrt{\pi}}\int_{-\infty}^\infty \exp^{\left(\frac{-v^2}{v_p^2}\right)}\rho_{12}\left(v,z\right) dv
\end{equation}
where $v_p=\sqrt{2k_B T/m}$ is the most probable velocity and we have explicitly noted that $\rho_{12}$ is a function of both $v$ and $z$. The double integration for the optically-thick, Doppler-averaged ensemble is performed numerically, where at each step $dz$ of the optically-thick integration, the entire ensemble must be integrated over all velocity classes. 

Given the importance of optical depth illustrated in the previous section, the effect of Doppler averaging on the OD alone is expected to cause significant reductions in the attainable sensitivity. To better illustrate this point, we allow our model to independently tune the Doppler width of the probing transition (with the full width at half maximum (FWHM) defined as $\Gamma_D=k_p v_p 2\sqrt{\ln2}$) without altering normally coupled model parameters such as the atom density. 

In Figure \ref{fig:DopplerLength}(a) we show the sensitivity as a function of total atom number for ensembles with $\Gamma_D=2\pi\times$\SIlist{1;3;10;25}{\mega\hertz} as the solid purple, dashed orange, dash-dot-dot green, and dotted red lines, respectively. For each point we have numerically optimized the Rabi frequencies to provide the optimum sensitivity (see App.~\ref{app:OptimalRabi}). As the Doppler width is increased, we see that the sensitivity in the linear $N$ regime worsens and the required number of total atoms to obtain the $\sqrt{N}$ regime is increased. Furthermore, the optimal Rabi frequencies necessary to obtain the $\sqrt{N}$ regime significantly increase with $\Gamma_D$ and generally satisfy $\Omega_p\approx\Omega_c\gg\Gamma_D$. For $\Gamma_D=2\pi\times\SI{25}{\mega\hertz}$, $\Omega_{p,c}^{opt}>2\pi\times\SI{130}{\mega\hertz}$ is necessary to obtain the $\sqrt{N}$ regime.

We also show the sensitivity of an ensemble with $\Gamma_D=2\pi\times\SI{514}{\mega\hertz}$, which corresponds to the typical D2 Doppler linewidth of rubidium, as the magenta line with diamond points and the cyan line with circular points. Given the prohibitive computational resources required to perform the double integration and the numerical optimization, only fixed Rabi frequencies are shown ($\Omega_{p,c}=2\pi\times\SI{15}{\mega\hertz}$ and \SI{40}{\mega\hertz} respectively). In this limit of fixed Rabi frequency well below $\Gamma_D$, we see that the sensitivity increases linearly with $N$ to a point where excessive absorption of the probing field, leading to increased measurement noise, becomes limiting.

In Figure \ref{fig:DopplerLength}(a) we also show the SQL result from Eq.~\ref{eq:SQL} as the gray dash-dot line which again assumes all atoms in the sensing volume for $N$ and a coherence time of $2/(2\gamma+\Gamma_r+\Gamma_c)$. We again find the ratio of the EIT sensitivity with that of the SQL, as shown in Figure \ref{fig:DopplerLength}(b). In the limit of sufficient optical depth and Rabi frequency, we note that all of the narrow Doppler width ensembles are able to obtain the same performance relative to the SQL as the Doppler-free case. 
For the full Doppler width of the magenta/diamond and cyan/circle traces, the performance relative to the SQL is significantly worse. This can be largely attributed to a reduction of the effective atom number (of order $\Gamma/\Gamma_D$) as atoms in high velocity classes do not participate in EIT. 

In Figure \ref{fig:DopplerLength}(c) we show the ratio of EIT sensitivity to the SQL as a function of $\Gamma_D$ at an $\OD=1$, which corresponds to taking a vertical slice of the values from part (b) near the optically-thin regime. Current state-of-the-art Rydberg sensor experiments using vapor cells often have $\OD\lesssim1$. They are also typically near room temperature and therefore most closely correspond to the upper-right portion of this figure (\emph{i.e.} low OD and large Doppler width). At $OD=1$ and $\Gamma_D=\SI{514}{\mega\hertz}$, the optimal sensitivity is \SI{36}{\deci\bel} worse than the SQL, or \SI{1.06}{\hertz/\sqrt{\hertz}}. Assuming an rf field measurement with a Rydberg sensor that has a discrimination coefficient equal to a dipole moment of $\wp=1574.9\,ea_0$ (corresponding to the $50D_{5/2}\rightarrow 51P_{3/2}$ Rydberg transition in Rubidium-85), this spectroscopic sensitivity yields a field sensitivity of \SI{8.4}{\nano\volt\per\meter\sqrt{\hertz}}.

We again note that the optical depth necessary to achieve the theoretical optimum relative to the SQL is generally high and increases substantially with $\Gamma_D$ (requiring values of $\OD>100$), and would be difficult in a thermal vapor. Furthermore, given that our simplified model does not include the quantum back-action that likely dominates at these extreme parameter regimes \cite{fleischhauer_quantum_2000} nor atom-atom interactions specific to Rydberg vapors that occur for high densities and optical powers \cite{weller_interplay_2019}, this level of performance cannot be guaranteed.
As a result, it is unlikely, even with unlimited optical powers, that a typical room-temperature ensemble will be able to approach the predicted SQL sensitivity.

\section{Discussion}

In summary, we have shown that in the optically-thin, Doppler-free, long-lived coherent state limit, the optimal ratio of probe to coupling Rabi frequencies is $\sqrt{2}$. We have numerically demonstrated that the performance of EIT probing relative to optimal SQL probing, of state perturbations due to external fields, is most strongly dependent on the optical depth (atom number) and the applied optical Rabi frequencies. EIT probing can approach the SQL only at high optical depths and high Rabi frequencies, and for an EIT-ladder scheme the closest approach is limited to approximately \SI{3.5}{\deci\bel} when perfectly optimized. This bound arises solely from absorption losses of the probe field and does not include other deleterious effects known to occur at high ODs and optical powers, such as back-action or radiation trapping of the probing field, which will further degrade the performance. Often OD and optical powers are limited for technical reasons and our results can be used to obtain the optimal performance of the EIT sensor in these realistic scenarios as well. We find that Doppler-broadening in room-temperature atoms reduces the effectiveness of EIT probing relative to the SQL by significant amounts (of order \SIrange{30}{40}{\deci\bel}). By combining these results, we can define a clear recipe for optimizing the sensitivity of Rydberg EIT experiments with varying degrees of optical depth and Doppler-averaging.

Our results show that while the potential sensitivity of standard quantum limited atomic sensors is attractive \cite{meyer_assessment_2020}, reaching that limit is challenging, especially with Doppler-broadened atoms. In this case, alternative probing schemes will likely be necessary to reach the SQL. This challenge is especially significant for current Rydberg electric field sensors, where higher densities and Rabi frequencies lead to increased dephasing from Rydberg-Rydberg collisions and large cells interfere with the RF radiation pattern to be detected. 

Doppler-free Rydberg excitation schemes using more than two optical fields \cite{carr_three-photon_2012,shaffer_read-out_2018,thaicharoen_electromagnetically_2019} are one path to achieve significant improvements in sensitivity, sidestepping the Doppler limits derived here.  Experimental implementation of Doppler-free Rydberg state spectroscopy in warm vapors have not yet demonstrated significant benefits to our knowledge.  One reason for this is that other broadening effects (such as transit broadening) can be equally limiting \cite{sibalic_dressed-state_2016}. It is also worth noting that Doppler-free excitation schemes are not equivalent to the zero-velocity treatment presented here, since a large distribution of single-photon detunings will remain. As a result, Doppler-free optical alignment schemes might require higher ODs and Rabi frequencies than the cold-atom, zero-velocity scheme.

An alternate approach to improving the capabilities of electric, or other field sensors, is to couple the atoms to classical apertures, waveguides, and/or resonators \cite{meyer_waveguide-coupled_2021,facon_sensitive_2016,suleymanzade_tunable_2020,morgan_coupling_2020}.  This allows one to increase the sensitivity by concentrating the RF mode size to the atomic sensing volume, effectively increasing the discrimination coefficient $d$. Ensembles of atoms, particularly laser-cooled atoms, are also amenable to large amounts of entanglement in the form of spin squeezing, that may also lead to significant future enhancements, reaching beyond the standard quantum limit to the Heisenberg limit.  

In any case, it is clear that atomic sensors are becoming useful tools to measure the electromagnetic spectrum for a wide array of applications.  More foundational research, to understand intrinsic capabilities and limitations of atomic quantum sensors and their readout mechanisms, will be critical to guide ongoing efforts.

\section*{Acknowledgments}

The authors thank Clement Wong for helpful discussions and recognize financial support from the Defense Advanced Research Projects Agency (DARPA) Quantum Apertures program. Dr.~O'Brien recognizes financial support from the Office of Naval Research (ONR) In-House Laboratory Independent Research (ILIR) program at the Naval Air Warfare Center Weapons Division. 

The views, opinions and/or findings expressed are those of the authors and should not be interpreted as representing the official views or policies of the Department of Defense or the U.S.~Government.

\appendix








\section{Model Definition} \label{app:Model}

\subsection{Hamiltonian}\label{app:H}

The level diagram for the typical Rydberg ladder-EIT three level system is shown in Figure \ref{fig:levelDiagram}. The corresponding Hamiltonian, in the rotating frame after discarding the counter-rotating terms, is
\begin{equation}\label{eq:HamLadder}
    H = \frac{\hbar}{2}
    \begin{pmatrix}
    0 & -\Omega_p & 0\\
    -\Omega_p & -2\delta_p & -\Omega_c\\
    0 & -\Omega_c & -2(\delta_p+\delta)
    \end{pmatrix}
\end{equation}
Note that this Hamiltonian maps to the more conventional lambda configuration typically considered in EIT by changing the sign of $\delta$.

\subsection{Master Equation}\label{app:MasterEq}

The Master equation that defines the equations of motion (or Optical Bloch Equations (OBEs)) for our atomic system is \cite{fleischhauer_electromagnetically_2005}
\begin{equation}
    \frac{d\rho}{dt} = \frac{-i}{\hbar}\left[H,\rho\right] + L
\end{equation}
where we have used the definition for the commutator and $L$ is the Lindbladian operator that defines the dephasing process present in the system.

The standard Lindbladian operator that defines the repopulation and relaxation terms of the master equation is defined as 
\begin{equation}
    \Gamma_{ij}\hat{L}_{ij}=\frac{\Gamma_{ij}}{2}\left(2\hat{\sigma}_{ji}\rho\hat{\sigma}_{ij}-\hat{\sigma}_{ii}\rho-\rho\hat{\sigma}_{ii}\right)
\end{equation}
where $\hat{\sigma}_{ij}=\ket{i}\bra{j}$ is the projection operator and $\Gamma_{ij}$ is the associated dephasing rate between levels $\ket{i}$ and $\ket{j}$. When $i\neq j$, this represents population transfer between states or a $T_1$ dephasing process. When $i=j$, this represents pure dephasing or a $T_2$ process. The Lindbladian accounting for natural lifetimes of the two excited states ($\Gamma$ and $\Gamma_r$) and collisional dephasing of the Rydberg state $\Gamma_c$, is written as
\begin{equation}\label{eq:LindOp}
    L=\Gamma\hat{L}_{21}+\Gamma_r\hat{L}_{32}+\Gamma_c\hat{L}_{33}
\end{equation}

Transit dephasing is another important type of dephasing common to thermal vapor sensors. It occurs as atoms transit the optical beams; entering in their ground state, interacting with the light fields, then leaving the light fields. As atoms leave, excited population and established coherence between states is lost and replaced by newly entering ground state atoms. The characteristic time for this process to occur is determined by the average time of flight through the cross section of the laser beams, and this defines the effective dephasing rate \cite{sagle_measurement_1996}.
\begin{equation}\label{eq:transit}
    \gamma=\sqrt{\frac{8 k_B T}{\pi m}}\frac{1}{w\sqrt{2\ln(2)}}
\end{equation}
where $w$ is the $1/e^2$ beam waist, $m$ is the atom mass, $T$ the ensemble temperature, and $k_B$ is Boltzmann's constant.

This process cannot be accurately captured by the Lindblad operator of Eq.~\ref{eq:LindOp} and is instead defined where the dephasing rates from all levels are given in the same form as in the Lindbladian above and the repopulation is limited to only the ground state and occurs at a fixed rate \cite{auzinsh_optically_2010}:
\begin{equation}\label{eq:Ltransit}
    L_\gamma = \gamma\hat{\sigma}_{11} - \frac{\gamma}{2}\left(I\rho +\rho I\right)
\end{equation}
where $I$ is the identity matrix spanning the Hilbert space and the repopulation is chosen to be limited to the ground state $\ket{1}$ in the case of ladder-EIT. 

\begin{figure}[b]
    \centering
    \includegraphics[width=\columnwidth]{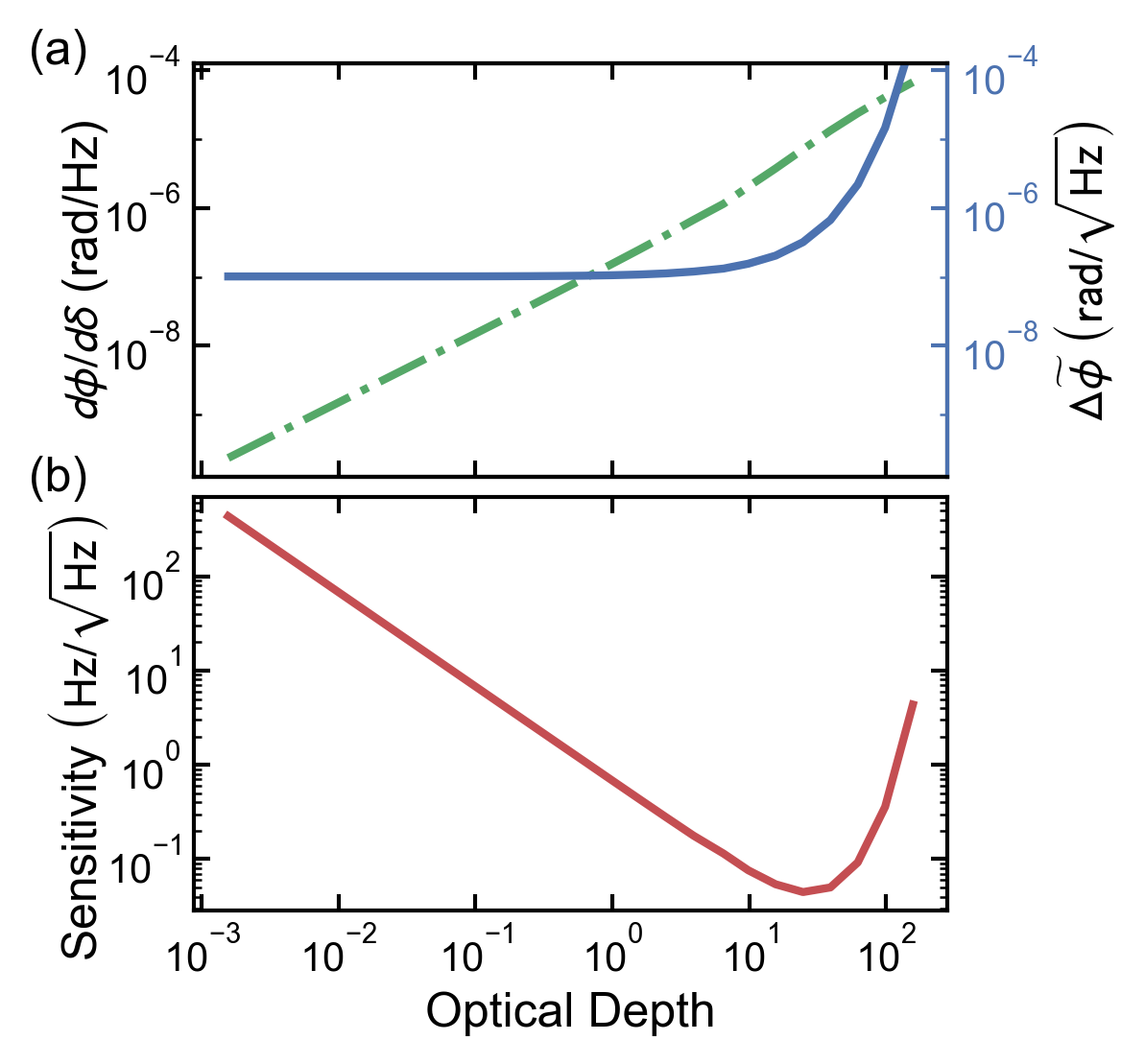}
    \caption{Slope, noise, and sensitivity for fixed optical Rabi frequencies versus optical depth.
    (a) Slope (dash-dot green) and associated photon shot noise (solid blue).
    (b) Sensitivity (red). Degraded sensitivity at large optical depth is due to increased phase noise in a progressively weaker transmitted probe.}
    \label{fig:OptThickSNR}
\end{figure}

\begin{figure*}[t]
    \centering
    \includegraphics[width=1.8\columnwidth]{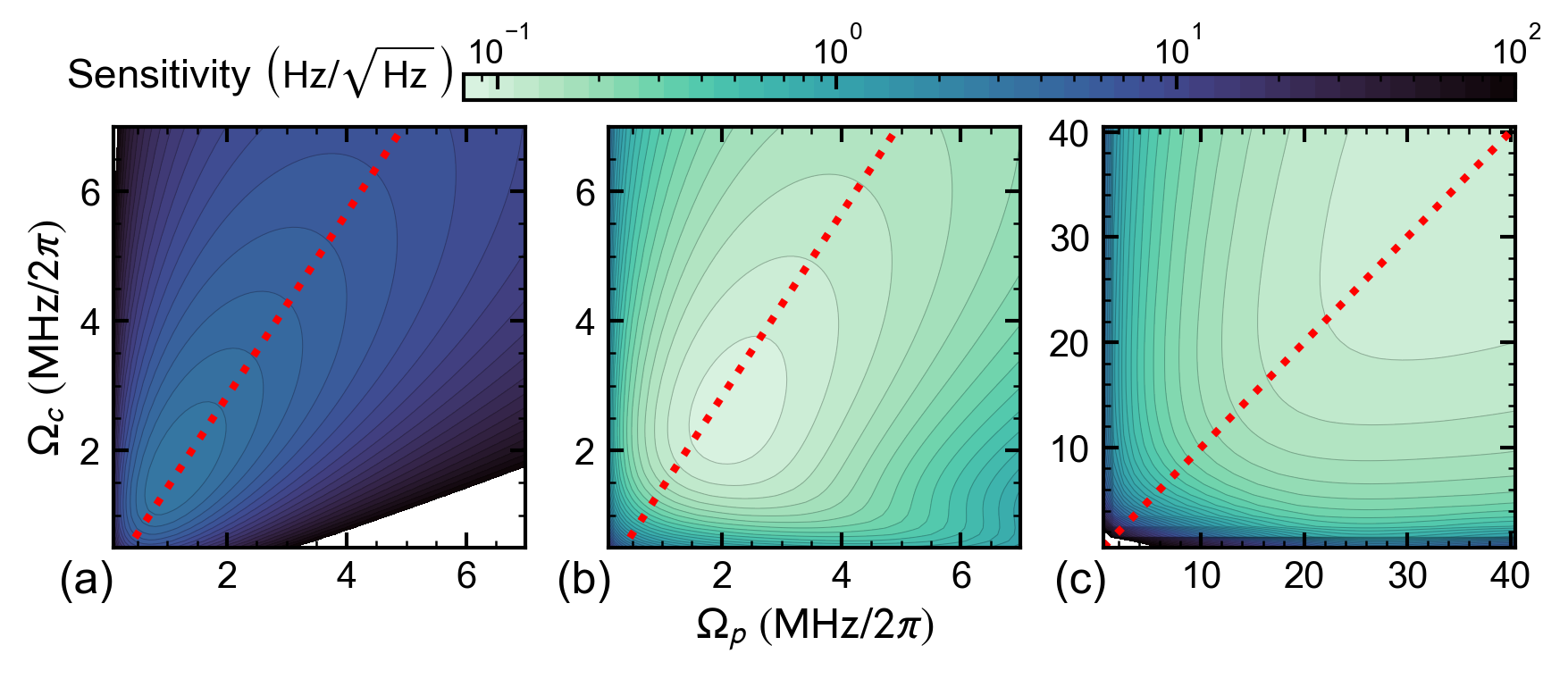}
    \caption{Sensitivity using EIT readout versus probe and coupling Rabi frequencies. Model parameters: $\Gamma=2\pi\times\SI{6.0666}{\mega\hertz}$, $\gamma=2\pi\times\SI{100}{\kilo\hertz}$, $\Gamma_r=2\pi\times\SI{1}{\kilo\hertz}$, $\Gamma_c=0$, $\delta_p=0$.
    (a) Optically-thin, Doppler-free regime: $\OD=0.16$. Red dashed line shows $\Omega_c=\sqrt{2}\Omega_p$.
    (b) Optically-thick, Doppler-free regime: $\OD=7.8$. Red dashed line shows $\Omega_c=\sqrt{2}\Omega_p$.
    (c) Optically-thick, Doppler-averaged regime: $\OD=12.5$. Red dashed line shows $\Omega_c=\Omega_p$.
    }
    \label{fig:Contours}
\end{figure*}

\subsubsection{Lambda-Configuration}
The above model definition is readily converted from a ladder-EIT system to a more typical lambda-EIT system where states $\ket{1}$ and $\ket{3}$ are ground states. First, the sign of $\delta$ is changed. Then the associated Lindbladian is altered to
\begin{equation}
    L=\frac{\Gamma}{2}\left(\hat{L}_{21}+\hat{L}_{23}\right) + \Gamma_c\left(\hat{L}_{33}+\hat{L}_{11}\right)
\end{equation}
which accounts for equal decay from the excited state to both ground states and both ground states are effected by some collisional broadening. For transit dephasing, we assume both ground states are thermally populated, so the projector $\hat{\sigma}_{11}$ in Eq.~\ref{eq:Ltransit} would be replaced with $(\hat{\sigma}_{11}+\hat{\sigma}_{33})/2$.

\subsection{Observables}\label{app:Observables}

In order to define the sensitivity limit using EIT readout, we must determine the effect of the atomic system on the probing light \cite{*[{See Chapter 10 of }] [{. Note this reference uses CGS units.}] auzinsh_optically_2010}. This is done by relating the polarizability of the atoms to the expectation value of the dipole operator $\vec{\mu}$
\begin{equation}
    \vec{P}=n\braket{\vec{\mu}}=\epsilon_0 \overleftrightarrow{\chi} \vec{E}(t,z)
\end{equation}
The above relationship is critical to relate the quantum dynamics to the observables (amplitude and phase) of the probe light field $E(t,z) = E(z) e^{-i(\omega t - \phi(z))}$.  We determine how the polarizability influences the transmitted light field, $E(t,z)$, using Maxwell's wave equation in a medium:
\begin{equation}
    \frac{\partial^2 E(t,z)}{\partial z^2}+k^2E(t,z) = -\frac{k^2}{\epsilon_0}P
\end{equation}

Assuming the probing field is linearly polarized and that changes in the transmitted field occur at lengths scales $\gg \lambda$, we can determine the first order differential equations for the transmitted probe field.  The dynamics of the probe amplitude $E(z)$ and phase $\phi(z)$ are given as (where we drop explicitly the $(z)$ dependence).
\begin{align}
    \frac{1}{E}\frac{dE}{dz} &= \frac{\omega}{2Ec\epsilon_0} \Im(P)=\frac{\omega n \mu^2}{2c\epsilon_0 \hbar \Omega_p} \Im(\rho_{12})\\
    \frac{d\phi}{dz} &= \frac{\omega}{2Ec\epsilon_0}\Re(P) = \frac{\omega n \mu^2}{2c\epsilon_0 \hbar \Omega_p} \Re(\rho_{12})
\end{align}
We group together the leading constants into a factor
\begin{equation}
    \kappa = \frac{\omega n \mu^2}{2c\epsilon_0 \hbar}
\end{equation}
and note that the normalized rate of change of the field amplitude is equal to the normalized rate of change of the probe Rabi frequency. This yields
\begin{align}
    \frac{d\Omega_p}{dz} &= \kappa \Im(\rho_{12})\\
    \frac{d\phi}{dz} &= \frac{\kappa}{\Omega_p} \Re(\rho_{12})
\end{align}
In the general case of an optically thick medium, both $\Omega_p$ and by extension $\rho_{12}$ are functions of $z$.

In the case of the lambda-EIT system, a third differential equation is necessary to account for absorption of the coupling field by atoms in the $\ket{3}$ ground state.
\begin{equation}
    \frac{d\Omega_c}{dz}=\kappa \Im(\rho_{32})
\end{equation}

\section{Determining Field Sensitivity}\label{app:FieldSens}

One can convert the sensitivities presented in this work to other fields (say a magnetic field) by applying the appropriate interaction moment $d$ that describes the interaction of the field $F$ with the highest excited state. The resulting interaction Hamiltonian $H_I=d\,F=\hbar\,\delta$ provides the conversion between $\delta_{min}$ and $F_{min}$. From Eq.~\ref{eq:SQL} of the main text we can define the standard quantum limited minimum detectable field as
\begin{equation}\label{eq:SQLfield}
    \delta F_{min} = \frac{\hbar}{d}\sqrt{\frac{2e}{N T_2 T}}
\end{equation}

In the case of Rydberg electrometry, which motivated this work, the appropriate interaction moment is the dipole moment for rf transitions between Rydberg states. These moments, when resonant, can be very large (exceeding $1000\,ea_0$) and lead to theoretical SQL field sensitivities on the order of \SI{10}{\pico\volt\per\centi\meter\sqrt{\hertz}} \cite{fan_atom_2015}. However, as we demonstrate in the main text, achieving the SQL using coherent spectroscopy of three-level atoms is not generally possible and becomes exceedingly difficult in small, Doppler-broadened ensembles.

\section{Optimal Optical Depth}\label{app:OptimalDepth}

In Figure \ref{fig:OptThickSNR} we show how the slope $d\phi/d\delta$, noise $\Delta\widetilde{\phi}$, and corresponding sensitivity $\Delta\widetilde{\phi}/(d\phi/d\delta)$ changes as a function of optical path length for fixed Rabi frequencies $\Omega_p=2\pi\times\SI{2.95}{\mega\hertz}$ and $\Omega_c=2\pi\times\SI{3.30}{\mega\hertz}$. At high optical depth, the slope does not appreciably change. However, the phase noise increases dramatically as the transmitted probe power is significantly attenuated. It is the corresponding increase in measurement phase noise that limits the attainable sensitivity.

\section{Optimal Rabi Frequencies}\label{app:OptimalRabi}

\begin{figure}[t]
    \centering
    \includegraphics{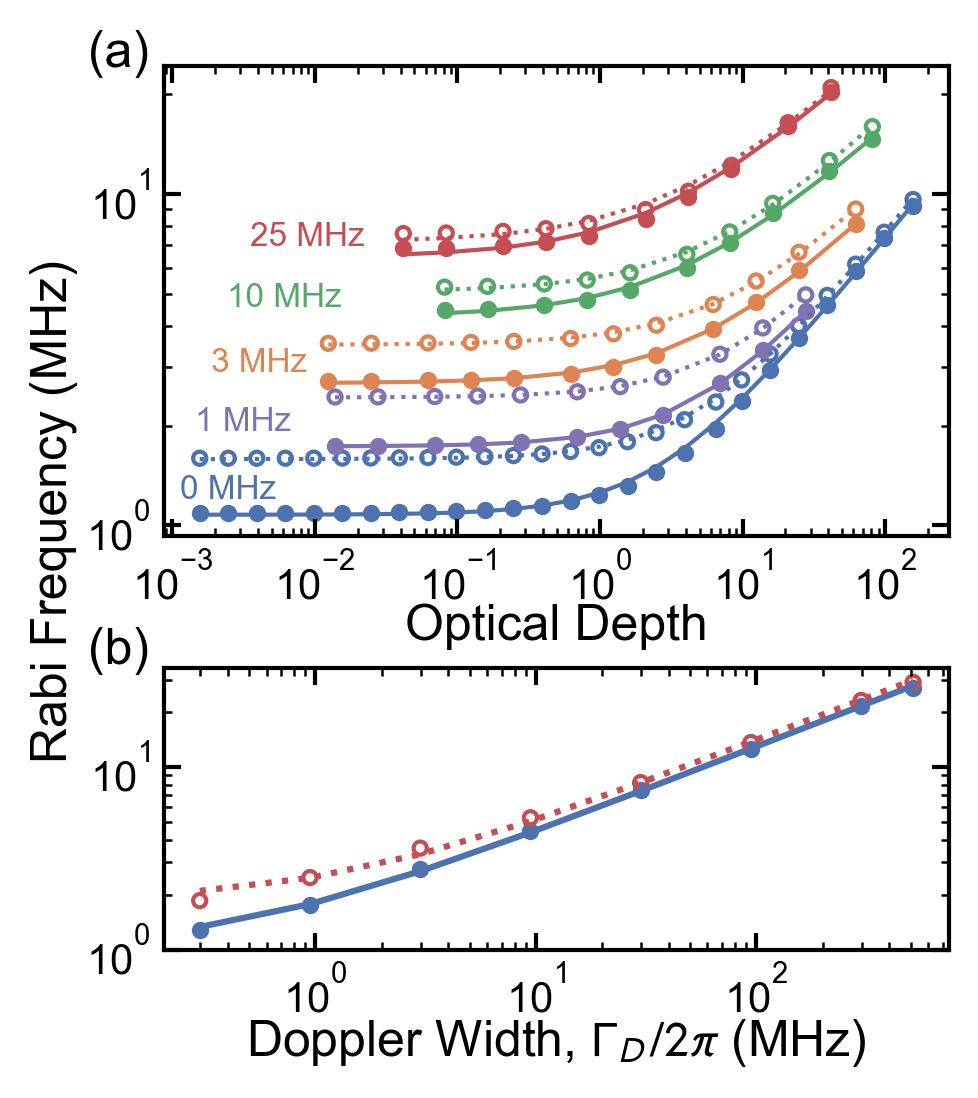}
    \caption{The optimized Rabi frequencies of ladder-EIT for the optimal sensitivity plots shown in Figs.~\ref{fig:OptThick} and \ref{fig:DopplerLength}.
    (a) The full (empty) points represent the optimal $\Omega_p$ ($\Omega_c$). The solid (dotted) line represents the empirical fit to the data. Each couplet of traces is labeled with the corresponding Doppler linewidth $\Gamma_D/2\pi$.
    (b) The solid blue (empty red) points show the optimal $\Omega_p$ ($\Omega_c$) versus Doppler width for an optically-thin sample. The solid (dotted) line shows the empirical fit.}
    \label{fig:OptThickRabis}
\end{figure}

Outside the optically-thin approximation, analytically determining the optimal probe and coupling Rabi frequencies is not generally possible and therefore requires numeric optimization techniques. These optimizations are equivalent to finding the minimum value of the sensitivity contours like those shown in Figure \ref{fig:Contours}.

The optically-thin, Doppler-free contour, for which the analytic results of the main text are approximately valid, is shown in Fig.~\ref{fig:Contours}(a). By visual inspection, it is clear that an optimum sensitivity exists at $\Omega_p=\sqrt{\Gamma(2\gamma+\Gamma_r+\Gamma_c)}=2\pi\times\SI{1.1}{\mega\hertz}$. It is also clear that this sensitivity lies on the line $\Omega_c=\sqrt{2}\Omega_p$ (shown as a dashed red line) and that this line approximates the principal plane of maximum curvature (\emph{i.e.} line for which the derivative along the perpendicular direction is equal to zero for all points).

An example of the deviation from the analytic results is shown in Figure \ref{fig:Contours}(b), where we show the sensitivity as a function of $\Omega_p$ and $\Omega_c$ at a fixed optical depth of 7.8. We have again overlaid the analytic $\Omega_c=\sqrt{2}\Omega_p$ result as a red dashed line. We can see that moving to the optically thick regime has moved the optimum away from the expected line and increased the necessary Rabi frequencies.

In Figure \ref{fig:Contours}(c) we show the contour plot for a Doppler-averaged, room-temperature ensemble with an optical depth of 12.56. We have also modified the overlaid trend line to be $\Omega_c=\Omega_p$ to match the empirically observed trend. 
As long as Doppler broadening is greater than the power broadening, there are no optimal Rabi frequencies and larger optical power yields better sensitivity. This is expected, because with larger power, more velocity classes participate in the measurement, up to the point that power broadening becomes greater than the Doppler width. More detailed models that account for additional effects may reveal optimal Rabi frequencies, but this is beyond the scope of this work.
For ladder-EIT schemes used with Rydberg state spectroscopy, the coupling transition typically has a small dipole moment and large Rabi frequencies are not readily obtainable.

\begin{table}[b]
    \centering
    \begin{tabular}{ccccc} \toprule
        & $\Gamma_D$ & $\Omega_0$ & $b$ & $m$ \\ \cmidrule(rl){2-5}
        & \si{\mega\hertz} & \si{\mega\hertz} & \si{\mega\hertz\squared} & \\ \midrule
      $\Omega_p$ & 0 & 1.08 & 0.65 & 0.527 \\
      $\Omega_c$ &  & 1.59 & 0.69 & 0.519 \\ \cmidrule(l){2-5}
      $\Omega_p$ & 1 & 1.73 & 0.78 & 0.499 \\
      $\Omega_c$ &  & 2.43 & 0.85 & 0.489 \\ \cmidrule(l){2-5}
      $\Omega_p$ & 3 & 2.66 & 1.30 & 0.422 \\
      $\Omega_c$ &  & 3.48 & 1.39 & 0.428 \\ \cmidrule(l){2-5}
      $\Omega_p$ & 10 & 4.27 & 2.51 & 0.394 \\
      $\Omega_c$ &  & 5.08 & 2.42 & 0.418 \\ \cmidrule(l){2-5}
      $\Omega_p$ & 25 & 6.46 & 4.45 & 0.390 \\
      $\Omega_c$ &  & 7.17 & 4.51 & 0.389 \\ \bottomrule
    \end{tabular}
    \caption{Fit values for the optimal Rabi frequencies for different Doppler widths.}
    \label{tab:FitVals}
\end{table}

In Figure \ref{fig:OptThickRabis} we show the optimal probe and coupling Rabi frequencies that were determined via numeric optimization and are used in Figures \ref{fig:OptThick} and \ref{fig:DopplerLength} of the main text. In the case of the Doppler-free, optically thick results, the sensitivity is optimized at every point (shown as the blue circles). For the Doppler-averaged results in Figure \ref{fig:DopplerLength}, numerical optimization is only done up to an optical depth of approximately 20. The optimized Rabi frequencies in each case are fit to an empirical function of the form
\begin{equation}
    \Omega_{opt}=\sqrt{\Omega_0^2+b \OD^m}
\end{equation}
where $\Omega_0$ corresponds to the asymptotic minimum optimum Rabi frequency and $m$, $b$ are empirical fit factors. The fit functions are then used to extrapolate the optimal Rabi frequencies when calculating the sensitivities of Figure \ref{fig:DopplerLength}. The fit values used are listed in Table \ref{tab:FitVals}. The corresponding fits are shown as the lines of Figure \ref{fig:OptThickRabis}(a). For all data sets we observe that the optimal Rabi frequencies approach the relation $\Omega_p=\Omega_c$ as the optical depth is increased.

Figure \ref{fig:OptThickRabis}(b) shows the numerically-optimized Rabi frequencies for an optically-thin sample with varying Doppler widths. We see that increasing $\Gamma_D$ results in optimal Rabi frequencies that again approach the $\Omega_p=\Omega_c$ relationship.


\bibliography{EITNoise}

\end{document}